\documentclass[12pt]{iopart}

\usepackage{epsf}
\usepackage{graphicx}

\begin{document}

\title[Deconfinement and Chiral Symmetry Restoration]{
Deconfinement and Chiral Symmetry Restoration}

\author{\'{A}gnes M\'{o}csy\dag\footnote[7]{Speaker at the conference.},
Francesco Sannino\ddag\ and Kimmo Tuominen\dag\dag}

\address{\dag\ Institut f\"ur Theoretische Physik, J.W.
Goethe-Universit\"at, Postfach 11 19 32, 60054 Frankfurt am Main, Germany}

\address{\ddag\ The Niels Bohr Institute, Blegdamsvej 17, DK-2100 Copenhagen
\O, Denmark}

\address{\dag\dag\ Department of Physics,
P.O. Box 35, FIN-40014 University of Jyv\"askyl\"a, Finland, and
Helsinki Institute of Physics, P.O. Box 64, FIN-00014 University
of Helsinki, Finland}

\begin{abstract}
We illustrate why color deconfines when chiral symmetry is
restored in gauge theories with quarks in the fundamental
representation, and while these transitions do not need to
coincide when quarks are in the adjoint representation,
entanglement between them is still present.
\end{abstract}

\vspace {-.5cm}
\paragraph{Introduction.}
One of the long-standing puzzles in theoretical physics is the
relation between confinement and chiral symmetry breaking.

In pure Yang-Mills theory the $Z_{N}$ center of the gauge $SU(N)$
group is a global symmetry. The Polyakov loop is a gauge invariant
operator charged under $Z_{N}$, whose expectation value vanishes
at low temperatures, and is nonzero above a critical temperature
$T_\chi$, when the center symmetry is spontaneously broken. This
feature, together with its relation to the infinitely heavy quark
free energy makes the Polyakov loop suitable to be the order
parameter for the deconfinement transition in pure gauge theory.
We denote with $\chi$ the polyakov loop field which is usually
denoted with $\ell$.

When quarks with finite masses are added to the theory in the
fundamental representation of the gauge group the $Z_N$ symmetry
is not exact. QCD with massless quarks exhibits chiral symmetry.
The order parameter, the chiral condensate, is zero above
$T_\sigma$, where chiral symmetry is restored. Here $\sigma$ is
the interpolating field associated to the scalar component of the
$\bar{q}q$ operator. For any finite quark mass chiral symmetry is
explicitly broken. When quarks are in the adjoint representation
the center group symmetry is intact. For realistic quark masses
there are no exact symmetries, 
but one can still follow the behavior of the condensates. Analysis
done on the lattice showed that with quarks in the fundamental
representation deconfinement (a rise in the Polyakov loop),
happens at the temperature where chiral symmetry is restored
(chiral condensate decreases) $T_\chi=T_\sigma$
\cite{Karsch:2001cy}. Lattice also revealed that when quarks are
in the adjoint representation deconfinement and the chiral
symmetry restoration do not happen at the same temperature,
$T_\sigma\simeq 8T_\chi$ \cite{Karsch:1998qj}. Despite the
attempts to explain these behaviors \cite{Brown:dm}, the
underlying reasons are still unknown. Lattice simulations for two
color QCD at non-zero baryon chemical potential observe
deconfinement for 2 color QCD and 8 continuum flavors. Here also,
the Polyakov loop rises when the chiral condensate vanishes, at
the same value of the chemical potential \cite{Alles:2002st}.

Our goal is to provide a simple unified way to describe all of
these features. We study the two color theory with $N_f$ flavors
in the chiral limit, since with only minor modifications of the
effective Lagrangian we can discuss the theory with quarks in the
fundamental and adjoint representation at nonzero temperature or
quark chemical potential. The results presented here are based on
our recent work \cite{Mocsy:2003qw} concerning the transfer of
critical properties from true order parameters to non-critical
fields, approach presented in \cite{Mocsy:2003tr,{Mocsy:2003un}},
envisioned first in \cite{Sannino:2002wb}. For a complete review
see \cite{Mocsy:2004yt}. The transfer of information is possible
due to the presence of a trilinear interaction between the light
order parameter and the heavy non-order parameter field, singlet
under the symmetries of the order parameter field. Due to this
interaction, the expectation value of the order parameter field in
the symmetry broken phase induces a variation in the expectation
value for the singlet field, and spatial correlators for the
non-critical fields are infrared dominated.

\begin{figure}[t]
\begin{center}
\includegraphics[width=11truecm, clip=true]{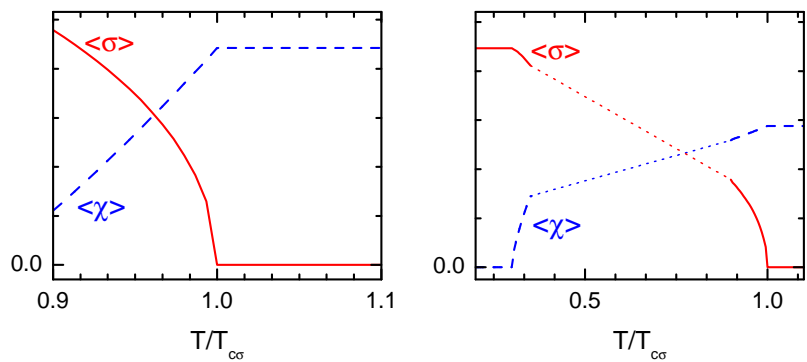}
\end{center}
\caption{Left panel: Expectation values of the Polyakov loop and
chiral condensate versus temperature, with massless quarks in the
fundamental representation. Right panel: Same as in left panel,
for the adjoint representation and $T_{\rm{c}\chi}\ll
T_{\rm{c}\sigma}$.} \label{Figura1}
\end{figure}

\vspace{-.3 cm}
\paragraph{Fundamental Representation.}
In two color QCD with two massless quark flavors in the
fundamental representation the global symmetry group is $SU(2N_f)$
which breaks to $Sp(2N_f)~$. The chiral degrees of freedom are
$2N_f^2-N_f-1$ Goldstone fields $\pi^a~$, and a scalar field
$\sigma~$, which is the order parameter. For $N_f=2$ the potential
is \cite{Appelquist:1999dq}:
\begin{eqnarray}
V_{\rm ch}[\sigma,\pi^a]&=&\frac{m^2}{2}{\rm Tr
}\left[M^{\dagger}M\right]+ {\lambda_1}{\rm Tr
}\left[M^{\dagger}M\right]^2+ \frac{\lambda_2}{4}{\rm Tr
}\left[M^{\dagger}MM^{\dagger}M\right] \label{chiralpot}
\end{eqnarray}
with $2\,M=\sigma + i\,2\sqrt{2}\pi^a\,X^a$, $a=1,\dots,5$ and the
generators $X^a\in {\mathcal A}(SU(4))-{\mathcal A}(Sp(4))$ are
provided explicitly in equation (A.5) and (A.6) of
\cite{Appelquist:1999dq}. The Polyakov loop, denoted by $\chi~$,
is treated as a heavy field singlet under the chiral symmetry. Its
contribution to the potential in the absence of the $Z_2$ symmetry
is
\begin{eqnarray}
V_\chi[\chi]=g_0\chi+\frac{m_\chi^2}{2}\chi^2+\frac{g_3}{3}\chi^3
+\frac{g_4}{4}\chi^4 \, . \label{chipot} \end{eqnarray}
The interaction terms allowed by chiral symmetry are
\begin{eqnarray}
V_{\rm{int}}[\chi,\sigma,\pi^a]&=& \left(g_1\chi
+g_2\chi^2\right){\rm Tr } \left[M^{\dagger}M\right]=\left(g_1\chi
+g_2\chi^2\right)(\sigma^2+\pi^a\pi^a) \, .
\end{eqnarray}
The $g_1$ term plays a fundamental role. In the symmetry broken
phase with $T<T_{c\sigma}$ the $\sigma$ acquires a non-zero
expectation value, which in turn induces a modification also for
$\langle\chi\rangle$. The extremum of the linearized potential,
near the phase transition where $\sigma$ is small, is at
\begin{eqnarray}
\langle\sigma\rangle^2 \simeq -\frac{m^2_{\sigma}}{\lambda}\, ,
\quad m^2_{\sigma}\simeq m^2 + 2g_1\langle\chi\rangle\, , \quad
\mbox{and} \quad \langle\chi\rangle\simeq -\frac{g_0}{m^2_{\chi}}
-\frac{g_1}{m_\chi^2}\langle\sigma\rangle^2\, ,
 \label{vevchi}
\end{eqnarray}
with $\lambda=\lambda_1 + \lambda_2$. Near $T_c$ the mass of the
order parameter field is assumed to posses the generic behavior
$m_\sigma^2\sim (T-T_{\rm{c}})^\nu$. For $g_1>0$ and $g_0<0$ the
expectation value of $\chi$ behaves oppositely to that of
$\sigma~$: As the chiral condensate starts to decrease towards
chiral symmetry restoration, the expectation value of the Polyakov
loop starts to increase, signaling the onset of deconfinement.
This is illustrated in the left panel of figure \ref{Figura1}.
When applying the analysis presented in
\cite{Mocsy:2003tr,{Mocsy:2003un}}, we find a drop near the
transitions, on both sides, in the spatial two-point correlator of
the Polyakov loop.

\vspace{-.3 cm}
\paragraph{Adjoint Representation.}
In two color QCD with two massless Dirac quark flavors in the
adjoint representation the global symmetry is $SU(2N_f)$ which
breaks via a bilinear quark condensate to $O(2N_f)$. There are
$2N_f^2+N_f-1$ Goldstone bosons, and two exact order parameter
fields, the chiral $\sigma$ field and the Polyakov loop $\chi$.
For $N_f=2$ the chiral part of the potential is given by
(\ref{chiralpot}) with $2\,M=\sigma + i\,2\sqrt{2}\pi^a\,X^a$,
$a=1,\dots,9$ and  the generators $X^a\in {\mathcal
A}(SU(4))-{\mathcal A}(O(4))$ are provided explicitly in equation
(A.3) and (A.5) of \cite{Appelquist:1999dq}. The $Z_2$ symmetric
potential for the Polyakov loop is
\begin{eqnarray}
V_\chi[\chi]=\frac{m_{0\chi}^2}{2}\chi^2+\frac{g_4}{4}\chi^4 \, ,
\end{eqnarray}
and the only interaction term allowed by symmetries is
\begin{equation}
V_{\rm{int}}[\chi,\sigma,\pi]=g_2\chi^2\,{\rm Tr
}\left[M^{\dagger}M\right]
 =g_2\chi^2(\sigma^2+\pi^a\pi^a) \, .
\end{equation}
Since the relevant interaction term $g_1\chi\sigma^2$ is now
forbidden, one might expect no information transfer between the
fields. Our analysis suggests though that this is not the case.
Consider the physical case in which the deconfinement happens
first \cite{Karsch:1998qj}, $T_{{\rm{c}}\chi}\ll
T_{{\rm{c}}\sigma}$. {}For $T_{{\rm{c}}\chi}<T<T_{{\rm{c}}\sigma}$
both symmetries are broken, and the expectation values of the two
order parameter fields are linked to each other:
\begin{eqnarray}
\langle\sigma\rangle^2=-\frac{m^2+
2g_2\langle\chi\rangle^2}{\lambda}\equiv
-\frac{m_\sigma^2}{\lambda}\, , \quad
\langle\chi\rangle^2=-\frac{m_{0\chi}^2+
2g_2\langle\sigma\rangle^2}{g_4}\equiv -\frac{m_\chi^2}{g_4}\, .
\label{vevad}
\end{eqnarray}
The behavior of $m_\chi^2\sim (T-T_{\rm{c}\chi})^{\nu_{\chi}}$ and
$m_\sigma^2\sim (T-T_{\rm{c}\sigma})^{\nu_{\sigma}}$ near
$T_{\rm{c}\chi}$ and $T_{\rm{c}\sigma}$, respectively, combined
with (\ref{vevad}), yields near these two transitions the
qualitative situation illustrated in the right panel of figure
\ref{Figura1}. On both sides of $T_{\rm{c}\chi}$
($T_{\rm{c}\sigma}$) the relevant interaction term $g_2\langle
\sigma\rangle\sigma\chi^2$ ($\langle \chi\rangle\chi\sigma^2$)
emerges, leading to the infrared sensitive contribution $\propto
\langle \sigma \rangle^2 /m_\chi~$
($\propto\langle\chi\rangle^2/m_\sigma$) to the $\sigma$ ($\chi$)
two-point function. Thus when $T_{\rm{c}\chi}\ll T_{\rm{c}\sigma}$
the two order parameter fields, a priori unrelated, do feel each
other near the respective phase transitions. The existence of
substructures near these transitions must be checked via lattice
calculations.

\vspace{-.3 cm}
\paragraph{Quark Chemical Potential.}
For two color QCD extending the discussion to finite chemical
potential is straightforward. With quarks in the pseudoreal
representation there is a phase transition from a quark-antiquark
condensate to a diquark condensate \cite{Hands:2001jn}. We hence
predict that when diquarks form for $\mu=m_{\pi}$, the Polyakov
loop feels the presence of the phase transition exactly in the
same manner as it feels when considering the temperature driven
phase transition. This was seen in lattice simulations
\cite{Alles:2002st}.

\vspace{-.3cm}
\paragraph{Discussion.}
Within an effective Lagrangian approach we have shown how
deconfinement (a rise in the Polyakov loop) is a consequence of
chiral symmetry restoration in the presence of massless quarks in
the fundamental representation. We expect this to hold for small
quark masses. If quark masses were very large then chiral symmetry
would be badly broken, and could not be used to characterize the
phase transition. But then $Z_N$ symmetry becomes more exact, and
its breaking would drive the (approximate) restoration of chiral
symmetry. In the non-perturbative regime the amount of $Z_N$
breaking is unknown, so we cannot establish which symmetry is more
broken for a given quark mass. We can make a rough estimate by
comparing two ratios: quark mass/confining scale, for chiral
symmetry breaking, and, unless some dynamical suppression,
$N_f/N$, for $Z_N$ breaking \cite{Mocsy:2004yt}. A two phase
transitions situation is still possible in QCD with massless
quarks in the limit of large number of colors for fixed number of
flavors, $N_f/N \ll 1~$, but this is unnatural in the case
$N_f\sim N$. Which of the underlying symmetries demands and which
amends can be determined directly from the critical behavior of
the spatial correlators of hadrons or of the Polyakov loop
\cite{Mocsy:2003tr,{Mocsy:2003un}}, and by extending our analysis
with the systematic study of the effects of quark masses, quark
flavors, anomalies, etc. This would help to further understand the
nature of phase transitions in QCD.\\


\begin{thebibliography}{0}

\bibitem{Karsch:2001cy}
F.~Karsch,
Lect.\ Notes Phys.\  {\bf 583}, 209 (2002)
[arXiv:hep-lat/0106019].

\bibitem{Karsch:1998qj}
F.~Karsch and M.~Lutgemeier,
Nucl.\ Phys.\ B {\bf 550}, 449 (1999) [arXiv:hep-lat/9812023].

\bibitem{Brown:dm} An incomplete list:
A.~M.~Polyakov,
Phys.\ Lett.\ B {\bf 72}, 477 (1978);
A.~Casher,
Phys.\ Lett.\ B {\bf 83}, 395 (1979);
C.~Adami, T.~Hatsuda and I.~Zahed,
Phys.\ Rev.\ D {\bf 43}, 921 (1991);
G.~E.~Brown {\it et al},
Nucl.\ Phys.\ A {\bf 560}, 1035 (1993); G.~E.~Brown {\it et al},
arXiv:hep-ph/0308147;
S.~Digal, E.~Laermann and H.~Satz,
Nucl.\ Phys.\ A {\bf 702}, 159 (2002); K.~Fukushima, in these
proceedings.

\bibitem{Alles:2002st}
B.~Alles, M.~D'Elia, M.~P.~Lombardo and M.~Pepe,
arXiv:hep-lat/0210039.

\bibitem{Mocsy:2003qw}
A.~Mocsy, F.~Sannino and K.~Tuominen,
accepted in PRL, arXiv:hep-ph/0308135.

\bibitem{Mocsy:2003tr}
A.~Mocsy, F.~Sannino and K.~Tuominen,
Phys.\ Rev.\ Lett.\  {\bf 91}, 092004 (2003)
[arXiv:hep-ph/0301229].

\bibitem{Mocsy:2003un}
A.~Mocsy, F.~Sannino and K.~Tuominen,
arXiv:hep-ph/0306069.

\bibitem{Sannino:2002wb}
F.~Sannino,
Phys.\ Rev.\ D {\bf 66}, 034013 (2002) [arXiv:hep-ph/0204174].

\bibitem{Mocsy:2004yt}
A.~Mocsy, F.~Sannino and K.~Tuominen,
arXiv:hep-ph/0401149.

\bibitem{Appelquist:1999dq}
T.~Appelquist, P.~S.~Rodrigues da Silva and F.~Sannino,
Phys.\ Rev.\ D {\bf 60}, 116007 (1999).

\bibitem{Hands:2001jn}
For a review on 2 color QCD see S.~Hands,
Nucl.\ Phys.\ Proc.\ Suppl.\  {\bf 106}, 142 (2002).

\end{thebibliography}
\end{document}